\documentclass[aps,prb,reprint,amsmath,amssymb,floatfix]{revtex4-2}
\usepackage{graphicx}
\usepackage[colorlinks, citecolor=blue, linkcolor=blue]{hyperref}
\usepackage{txfonts}
\usepackage{bm}		% Bold math

\newcommand{\CNN}{Centre de Nanosciences et de Nanotechnologies, CNRS, Universit\'e Paris-Saclay, 91120 Palaiseau, France}

\begin{document}

\title{Excitation of vortex core gyration in nanopillars through driven Floquet magnons}

\author{Gauthier Philippe}
\email{gauthier.philippe@universite-paris-saclay.fr}
\author{Joo-Von Kim}
\email{joo-von.kim@universite-paris-saclay.fr}
\affiliation{\CNN}

\begin{abstract}
The dynamics of vortex states in confined geometries like thin-film disks are characterized by a sub-GHz gyration, representing the damped oscillatory motion of the vortex core about the disk center. It has recently been shown that interactions between the core and azimuthal spin waves, lying in the GHz range and driven by magnetic fields, can result in steady-state core gyration. The gyration in turn provides a time-periodic modulation for the spin waves, resulting in the emergence of Floquet states. Here, we present results of a theoretical and computational study in which we examine how Floquet modes sustain this core gyration. In particular, we find that multiple steady-state gyration radii are possible under certain field conditions, resulting from the nonlinear interactions between the core and Floquet modes. Different gyration radii result in distinct Floquet frequency comb spectra and allow for hysteretic effects, as reported in recent experiments.
\end{abstract}

\date{\today}

\maketitle

%%
%	Section: Introduction
%%
\section{Introduction \label{introduction}}
The magnetic vortex state in ferromagnetic films represents a curling configuration of the magnetic moments within the film plane, culminating with a magnetization perpendicular to the plane at its center within a region called the vortex core. The vortex is an equilibrium ground state in thin ferromagnetic disks of certain aspect ratios~\cite{guslienko2001evolution, metlov2002stability, scholz2003transition, guslienko2004vortex}. Two distinct categories of excitations exist about this ground state. The first involves the translational motion of the vortex core around the disk center, termed the gyrotropic mode. The lowest order gyrotropic mode typically lies in the sub-GHz range~\cite{guslienko2002eigenfrequencies, ivanov2004gyrotropic}, while higher-orders involving standing wave profiles along the film thickness can reach tens of GHz~\cite{ding2014higher, bondarenko2024dominant}. The second are spin wave modes, which represent fluctuations about the static vortex background. These modes typically lie in the GHz range and are geometrically quantized; they reflect the axial symmetry of the vortex ground state and are well-described by radial and azimuthal indices $(n,m)$ to lowest order~\cite{novosad2002spin, ivanov2002magnon, park2003imaging, buess2005excitations, ivanov2005high, awad2010precise, castel2012perpendicular, taurel2016complete, schultheiss2019excitation}. Due to interactions with the vortex core, it is known that there is a lifting of the degeneracy between clockwise and counterclockwise propagation of these spin waves, resulting from dynamical effects~\cite{park2005interactions, guslienko2008dynamic} or dipolar interactions~\cite{koerber2022mode, uzunova2023nontrivial, salama2025large}.

Within this context, it is interesting to enquire how vortex core motion, i.e., gyration about the disk center, affects the spin waves. For a static displacement of the vortex core from the disk center, experiments and simulations have already shown that the breaking of the axial symmetry has a strong influence on the normal modes~\cite{koerber2023modification}. One way to account for a dynamical core involves introducing an additional interaction with a topological gauge field that arises from the moving background~\cite{guslienko2010topological}. Moreover, because steady-state gyration represents a time-periodic motion of this background, a fruitful line of inquiry has involved viewing the problem through the lens of Floquet theory~\cite{oka2019floquet, rodriquezvega2021low}. As recent work has shown, driving the core gyration with a radiofrequency (RF) magnetic field results in the appearance of Floquet magnon states, which manifest themselves as a rich frequency comb with a frequency spacing dictated by the gyration frequency~\cite{heins2024self}. More strikingly, strong field pumping of azimuthal spin wave modes can also trigger steady-state gyration, representing a ``self-induced'' mechanism for generating Floquet states. Subsequent work has highlighted the key role of the vortex core, whose removal through physical patterning results in the suppression of Floquet states~\cite{heins2025control}.

In this article, we discuss a theoretical and computational study in which we examine how interactions between azimuthal spin wave modes and the vortex core can result in self-sustained gyration of the core in a nanopillar geometry. A key finding is that certain field driving amplitudes and frequencies can result in multiple metastable gyration radii, which result in distinct frequency comb spectra. This metastability implies that the self-induced Floquet mechanism can be hysteretic, highlighting the strong nonlinear processes that underpin the phenomenon.

%%
%	Section: Geometry and method
%%
\section{Geometry and method \label{sec:structure}}
We studied the magnetization dynamics within ferromagnetic thin-film disk with a diameter of $300$~nm and a thickness of $20$~nm, which is representative of a free magnetic layer in a magnetic tunnel junction~\cite{martins2023second}. The magnetization dynamics was investigated within the micromagnetics approximation. We solved numerically the Landau-Lifshitz equation with Gilbert damping (LLG),
\begin{equation}
\frac{d\mathbf{m}}{dt} = -|\gamma_0| \mathbf{m} \times \mathbf{H}_\mathrm{eff} + \alpha \mathbf{m} \times \frac{d\mathbf{m}}{dt},
\label{eq:LLG}
\end{equation}
where $\mathbf{m} = \mathbf{m}(\mathbf{r},t)$ is a unit vector representing the magnetization field, $\gamma_0 = \mu_0\gamma$ is the gyromagnetic constant, and $\alpha$ is the damping constant. The effective field,  $\mathbf{H}_\mathrm{eff} = -(1/\mu_0 M_s) \delta U/\delta \mathbf{m}$, represents the variational derivative of the magnetic energy $U$ with respect to the magnetization. Contributions to $U$ considered are the exchange, dipolar, and Zeeman interactions. We assumed a saturation magnetization of $M_s = 1.087$~MA/m, an exchange constant of $A = 18$~pJ/m, and a Gilbert damping constant of $\alpha = 0.008$. This is consistent with typical CoFeB-based alloys used magnetic tunnel junctions. With these geometric and material parameters, the ground state configuration is a vortex state. Equation~(\ref{eq:LLG}) was solved with the finite different scheme, using 128$\times$128$\times$1 finite difference cells to discretize the disk. Two open-source codes were used. First, the \texttt{magnum.np} code~\cite{bruckner2023magnum} was employed to compute the spin wave eigenmodes about the static vortex ground state, using the built-in routine that solves the linearized LLG equation. %The spatial profiles obtained of the eigenmodes subsequent serve as a basis for the mode-projection method~\cite{massouras2024mode}. 
Second, time integration of the full LLG equation was performed using the \texttt{mumax3} code~\cite{vansteenkiste2014design}, which provides the full time evolution of the magnetization within each finite difference cell. Coarse-graining of this output using the supercell method~\cite{massouras2024mode} allowed as to obtain accurate estimates of the power spectral density of excitations and spatial profiles of nonequilibrium modes without requiring significant postprocessing.

%%
%	Section: Results
%%
\section{Results}
%%
%%	Sub-section
\subsection{Frequency comb and gyration radius}

\begin{figure} 
	\centering
	\includegraphics[width=8.5cm]{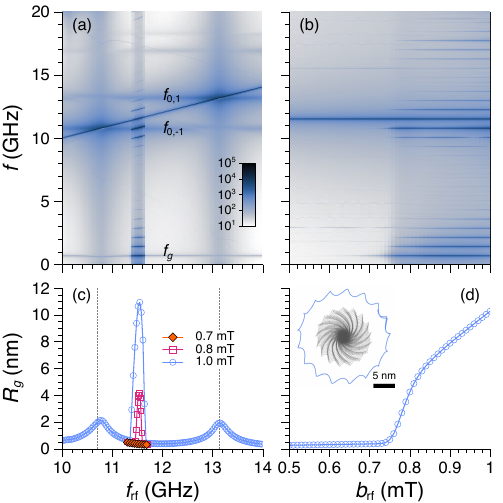}
	\caption{Color map of the power spectral density $S(f)$ as a function of the (a) excitation frequency, $f_\mathrm{rf}$, for $b_\mathrm{rf} = 1$~mT and the (b) excitation field, $b_\mathrm{rf}$, for $f_\mathrm{rf} = 11.5$~GHz. (c) Gyration radius of the vortex core, $R_g$, as a function of $f_\mathrm{rf}$. Dashed vertical lines indicate the frequencies of the linear modes $m=\pm 1$. (d) $R_g$ as a function of $b_\mathrm{rf}$. The inset shows the vortex core trajectory at $b_\mathrm{rf} = $ 1~mT, with the gray curve representing the transient dynamics over the first 100 ns of the simulation and the blue curve over 1.5 ns (i.e., just over a gyration period) at steady state. }
	\label{fig:FreqComb}
\end{figure}

We first examine the power spectrum of spin wave excitations as a function of a driving RF field, $\mathbf{b}_\mathrm{rf} = b_\mathrm{rf} \cos{(2\pi f_\mathrm{rf} t) \hat{\mathbf{x}} } $, applied along the $x$ direction in the film plane. The resulting power spectrum for $b_\mathrm{rf} = 1$~mT as a function of the driving frequency $f_\mathrm{rf}$ is shown in Fig.~\ref{fig:FreqComb}(a).
For each value of $f_\mathrm{rf}$, we simulated the dynamics over an interval of $t_0=$ 100~ns at zero temperature using the equilibrium vortex state as the initial state. The power spectral density (PSD) [$S(f)$] shown is computed as follows. The perpendicular magnetization component $M_z(\mathbf{r},t)$ is spatially averaged over each supercell $j$, comprising $4 \times 4$ finite difference cells, and recorded as a function of time; call this quantity $M_j(t)$. The power spectral density within each supercell, $S_{j}(f)$, is computed from the square of the discrete Fourier transform of $M_j(t)$ with the Hann windowing function $w(t)$, and $S(f)$ is then obtained by averaging over the $N$ supercells,
\begin{equation}
S(f) = \frac{1}{N} \sum_{j=1}^{N} S_j(f) = \frac{1}{N} \sum_{j=1}^{N} \left| \int_{0}^{t_0} dt \, w(t)\, e^{-i 2\pi f t} \, M_j(t) \right|^2 .
\end{equation}

Figure~\ref{fig:FreqComb}(a) exhibits a number of features that highlight both the linear eigenmodes of the system and their evolution under the RF drive, represented by the diagonal line $f = f_\mathrm{rf}$. First, a number of horizontal lines can be seen. The lowest, at $f = f_g \approx 727$~MHz, corresponds to the frequency of vortex core gyration. The lines at 10.8 and 13.2 GHz correspond to the linear frequencies $f_{n,m}$ of the azimuthal modes $(n=0,m=\pm 1)$, while the lines at 17 and 18 GHz correspond to the $(n=1,m=\pm 1)$ modes. These are consistent with the results obtained using the eigenmode solver of \texttt{magnum.np} with the same micromagnetic parameters. When the drive frequency matches the linear frequencies of either of the two $(n=0,m=\pm 1)$ modes, $f_{0,\pm 1}$, we observe a strong response in the PSD as expected. More interestingly, however, a frequency comb appears for an interval of drive frequencies in between the two mode frequencies $f_{0,\pm 1}$. The comb spacing is determined by the gyration frequency, so we can write the comb frequencies as $f = f_\mathrm{rf} + k f_g$ with $k$ being an integer. We can also observe a large number of harmonics of $f_g$ at low frequencies within this interval.

As observed in larger disks~\cite{heins2024self}, the frequency combs only appear above a finite threshold of the applied RF field. This behavior is also seen in our system, as displayed in Fig.~\ref{fig:FreqComb}(b), which shows a color map of the PSD as a function of $b_\mathrm{rf}$ at a fixed frequency of $f_\mathrm{rf}=11.5$~GHz. Below $b_\mathrm{rf} \approx 0.75$~mT, a single intense peak at $f_\mathrm{rf}$ is present, corresponding to the direct response to the driving field. As $b_\mathrm{rf}$ is increased, a frequency comb centered around $f_\mathrm{rf}$ emerges, accompanied by a strong response in the gyration frequency and its harmonics. The number of sidebands also increase as a function of $f_\mathrm{rf}$.

The number and intensity of sidebands in the frequency comb increase as the average gyration radius of the vortex core increases. This behavior is displayed in Fig.~\ref{fig:FreqComb}(c) where the gyration radius $R_g$, averaged over an interval of 200~ns in the steady-state regime, is shown as a function of $f_\mathrm{rf}$ for three values of $b_\mathrm{rf}$. For a sub-threshold excitation, $b_\mathrm{rf}=0.7$~mT, the gyration radius remains close to zero as the frequency is swept across the region of interest. As the driving field is increased above the threshold of $b_\mathrm{rf} \approx 0.75$~mT, a peak in $f_g$ is observed around $f_\mathrm{rf} = 11.5$~GHz, which grows in amplitude as $b_\mathrm{rf}$ is increased. For the largest value of $b_\mathrm{rf}=1.0$~mT considered, two broader peaks in $T_g$ can also be seen, as indicated by the dashed lines Fig.~\ref{fig:FreqComb}(c). The positions of these peaks correspond to the linear frequencies of the $f_{0,\pm 1}$ modes. This result shows that driving the core gyration using high-frequency excitations is most efficient for the self-induced Floquet mode, rather than the linear azimuthal $(0,\pm 1)$ modes.

The threshold behavior for the self-induced excitation of Floquet magnons is further exemplified in Fig.~\ref{fig:FreqComb}(d), where $R_g$ is shown as a function of $b_\mathrm{rf}$ for $f_\mathrm{rf}=11.5$~GHz. We estimate the threshold to be $b_\mathrm{rf} \approx 0.75$~mT from this curve. The inset illustrates the vortex core trajectory under an applied field of $b_\mathrm{rf}=$ 1.0~mT for two different regimes. The gray curve shows the transient dynamics as the core is excited from its equilibrium position, over an interval of 100 ns. In contrast to a regular spiral growth, which is expected when the core gyration is excited directly, the trajectory exhibits instead a trochoidal motion~\cite{ivanov1998magnon, ivanov2010nonnewtonian}, resulting from interactions with the driven azimuthal modes. The blue curve corresponds to the trajectory at steady state over 1.5 ns, which is just over one period of gyration. Again, the trajectory deviates from the circular form expected with direct excitation, with approximately 17 cusps appearing over one period of gyration.

\subsection{Mode profiles and Floquet spectra}

\begin{figure} 
	\centering
	\includegraphics[width=8.5cm]{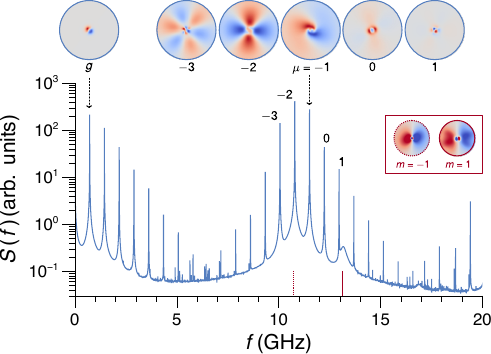}
	\caption{Power spectral density $S(f)$ of spin wave modes excited under $f_\mathrm{rf} = 11.5$~GHz and $b_\mathrm{rf}= 1.0$~ mT. The inset above the graph shows the spatial profiles of selected Floquet modes. The middle, right inset shows the spatial profiles of the linear $(n,m) = (0,\pm 1)$ modes, with the red vertical lines at 10.7 and 13.1 GHz indicating their frequencies.}
	\label{fig:SpectrProfil}
\end{figure}

We now turn our attention to the spatial profiles and power spectra of the Floquet modes, induced by direct excitation of the high-frequency azimuthal modes. The PSD $S(f)$ is shown in Fig.~\ref{fig:SpectrProfil} for $f_\mathrm{rf}=11.5$~GHz and $b_\mathrm{rf}=1.0$~mT.
The figure highlights clearly the two sets of frequency combs present, the first centered about the excitation frequency and the second comprising the harmonics of the gyration frequency. The vertical lines indicate the positions of the linear azimuthal mode frequencies $f_{0,\pm 1}$, with the corresponding spatial profiles shown in the middle inset. The top inset above the graph shows the extracted spatial mode profiles of the Floquet modes from the supercell method. We observe that the directly excited mode is indeed a mode with azimuthal symmetry $m=1$, however its profile is distorted with respect to the linear eigenmode. This difference has been observed in micron-sized permalloy disks and underlies the fact that the driven modes are Floquet modes~\cite{heins2024self}. Within the main comb, the azimuthal number changes by an increment of one as we move from the central peak, as shown in the inset. This can be understood within a particle scattering picture whereby the gyrotropic mode carries an azimuthal number of one. The Floquet modes within the comb, which we write symbolically as $| \mu \rangle$ (in order to distinguish them from the linear eigenmodes, $| m \rangle$), can then be thought of as arising from three-particle splitting and confluence processes of the form $|\mu\pm1\rangle \leftrightarrow |\mu\rangle \pm |g\rangle$. The comb modes are generated from the state $| \mu  = 1 \rangle$, which evolves from the $| m =1\rangle$ state that is initially populated by the external field. These scattering processes conserve both the azimuthal number (angular momentum) and the frequency (energy), $f_{\mu\pm1}=f_\mu \pm f_g$. We will further examine the consequences of the nonlinear interactions behind these scattering processes in the next section.

\begin{figure}
	\centering
	\includegraphics[width=8.5cm]{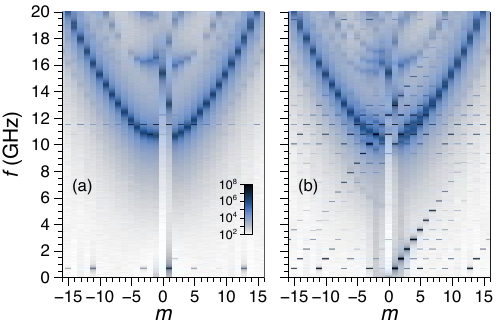}
	\caption{Color map of the azimuthal number-resolved power spectral density, $S_m(f)$, under an excitation frequency of $f_\mathrm{rf} = 11.5$~GHz and an excitation amplitude of (a) $b_\mathrm{rf} = 0.5$~mT and (b) $b_\mathrm{rf} = 1.0$~mT.}
	\label{fig:Modifl}
\end{figure}

We examine these scattering processes under a different light in Fig.~\ref{fig:Modifl}, where we present the azimuthal number-resolved power spectral density for two different excitation field strengths at a driving frequency of $f_\mathrm{rf} = 11.5$~GHz.
In contrast to the previous examples in which we relied on supercells to compute the power spectra, here we use the azimuthal function $\exp(i m \varphi)$, where $\varphi$ is the azimuthal coordinate in cylindrical coordinates, as a spatial filter to project out the corresponding Fourier coefficients. More precisely, we decompose the time evolution of the total magnetization computed, $\mathbf{M}(\mathbf{r},t) = \mathbf{M}_0(\mathbf{r}) + \delta \mathbf{M}(\mathbf{r},t)$, into a static part (representing the vortex ground state) $\mathbf{M}_0(\mathbf{r})$ and a dynamic part $\delta \mathbf{M}(\mathbf{r},t)$, and analyse the dynamics of the latter in terms of the Fourier components $a_m(t)$,
\begin{equation}
a_m(t) = \frac{1}{V} \int dV \,  \left( \delta \mathbf{M}(\mathbf{r},t) \cdot \hat{\mathbf{n}}(\mathbf{r}) \right) e^{i m \varphi},
\end{equation}
with $\hat{\mathbf{n}}(\mathbf{r}) = \hat{\mathbf{e}}_\rho \times \mathbf{M}_0(\mathbf{r})$ representing one of the transverse directions to $\mathbf{M}_0(\mathbf{r})$ ($\hat{\mathbf{n}} \approx \hat{\mathbf{e}}_z$ everywhere except within the vortex core). The color map in Fig.~\ref{fig:Modifl} therefore represents the $m$-resolved power spectral density, 
\begin{equation}
S_m(f) = \left| \int_{0}^{t_0} a_m(t) e^{-i 2 \pi f t} \right|^2.
\end{equation}
Moreover, we assumed a finite temperature of 1~K to thermally-populate all spin wave modes.
Below the threshold for a driving field amplitude of $b_\mathrm{rf} = 0.5$~mT [Fig.~\ref{fig:Modifl}(a)], we observe that the main response is at $f_\mathrm{rf} = 11.5$~GHz and $m=\pm 1$, along with other odd harmonics in $m$, by virtue of the fact that we use a linearly polarized excitation field. We can also discern the dispersion relation of the $n=0$ branch, given by a quadratic-like function in $m$ with a gap of around 10.9 GHz, and the bottom of the $n=1$ branch, whose minimum is around 16 GHz. In contrast to the zero temperature case considered above, thermal fluctuations here result in a finite gyration radius, resulting in a sizeable response in Fig.~\ref{fig:Modifl}(a) at $m=1$.

When the driving field exceeds the threshold for the self-induced Floquet mechanism, such as for $b_\mathrm{rf} = 1.0$~mT [Fig.~\ref{fig:Modifl}(b)], we observe the appearance of a checkerboard-like pattern, defined broadly by a set of parallel lines of the form $f = f_g (m-m_0)$ where $m_0$ is an integer offset. These lines represent a decomposition in $m$ of the frequency combs seen in Fig.~\ref{fig:SpectrProfil}. Each line can be interpreted as a sequence of scattering events given by $|\mu\pm1\rangle \leftrightarrow |\mu\rangle \pm |g\rangle$, providing a visual description of how the initial driven mode at 11.5 GHz populates the other Floquet modes within the frequency comb.

\subsection{Gyration due to magnon-core scattering}
We can gain insight into how the population dynamics of the Floquet modes sustain the steady gyration of the vortex core from the perspective of the Thiele model~\cite{thiele1973steady}. The Thiele model is based on the assumption that the magnetization dynamics is captured entirely by the translational motion of the vortex core. Let $\Theta = \Theta(\mathbf{r},t)$ and $\Phi = \Phi(\mathbf{r},t)$ represent the orientation of the magnetization field in spherical coordinates, $\mathbf{M}(\mathbf{r},t) = \left( \cos{\Phi} \sin{\Theta}, \sin{\Phi} \sin{\Theta}, \cos{\Theta}  \right)$. With the rigid-core \emph{ansatz}, $\Theta=\Theta_0[\mathbf{r}-\mathbf{X}(t)]$ and $\Phi=\Phi_0[\mathbf{r}-\mathbf{X}(t)]$, where $[\Theta_0(\mathbf{r}), \Phi_0(\mathbf{r})]$ is the static, equilibrium vortex configuration and by considering the core position within the film plane, $\mathbf{X}(t) = [X(t),Y(t)]$, as the sole dynamical variable, we arrive at the familiar form of the Thiele equation after integrating over all space,
\begin{equation}
\mathbf{G} \times \dot{\mathbf{X}} + \alpha \mathbf{D} \cdot \dot{\mathbf{X}} = -\frac{\partial U}{\partial \mathbf{X}},
\label{eq:Thiele}
\end{equation}
where $\dot{\mathbf{X}} =  d\mathbf{X}/d t$ denotes a time derivative. The first term on the left-hand side represents the gyrotropic term, where $\mathbf{G} = \int dV \left( \nabla\Theta_0 \times \nabla\Phi_0 \right) \sin\Theta_0$. The second term on the left-hand side represents Gilbert damping, with $\alpha$ being the damping constant and $\mathbf{D}$ is the damping dyadic, $\mathbf{D} = \int dV \left[ \nabla\Theta_0 \otimes \nabla\Theta_0 + \sin^2\Theta_0 \left(\nabla\Phi_0 \otimes \nabla\Phi_0 \right) \right] $, where $\otimes$ represents a direct product. $\mathbf{D}$ generally only contains diagonal terms and so it can be treated as a vector, i.e. $\mathbf{D} = D (1,1)$. The term on the right-hand side represents a force acting on the core position, where $U$ is a magnetic energy encompassing the exchange, dipolar, and Zeeman interactions. To a good approximation, $U = \kappa_1 \| \mathbf{X}\|^2 + \kappa_2 \| \mathbf{X}\|^4$, representing the confining potential for the vortex. The stiffness constants $\kappa_{1,2}$ are determined by $M_s$ and the geometrical parameters of the thin-film disk~\cite{ivanov2007excitation}.

For what follows, it will be convenient to express the Thiele equation [Eq. (\ref{eq:Thiele})] using polar coordinates $(R,\Psi)$, i.e., $R = \sqrt{X^2+Y^2}$ and $\Psi=\arctan{Y/X}$. Thus, 
\begin{subequations}
\begin{align}
\dot{R} &= -\left(\frac{\alpha D}{G^2 + \alpha^2 D^2 } \right) \frac{\partial U}{\partial R}, \\
\dot{\Psi} &= \left(\frac{G}{G^2 + \alpha^2 D^2}\right)\frac{1}{R} \frac{\partial U}{\partial R} \equiv \omega_g,
\end{align}\label{eq:ThielePolar}
\end{subequations}
with $\omega_g = 2 \pi f_g$ being the gyration frequency. With $\kappa \equiv \alpha D/G$, we can rearrange Eq.~(\ref{eq:ThielePolar}) to simplify the expression for the radial dynamics,
\begin{equation}
\dot{R} = - \kappa \omega_g R \equiv - \Gamma_{g} R,
\end{equation}
where $\Gamma_g$ represents the intrinsic Gilbert damping rate of the gyrotropic mode. For the system studied, we find $\Gamma_g = 5.86 \times 10^7$~rad/s. We note that the polar-coordinate form of the Thiele equation [Eq.~(\ref{eq:ThielePolar})] can be derived from the Euler-Lagrange equations, e.g., for $R$,
\begin{equation}
\frac{d}{d t}\frac{\partial L}{\partial \dot{R}} - \frac{\partial L}{\partial R} + \frac{\partial W}{\partial \dot{R}} = 0,
\end{equation}
with an analogous expression for $\Psi$. The corresponding Lagrangian is given by
\begin{equation}
L = L_B - U = \frac{1}{2}G \, R^2 \dot{\Psi}-U(R),
\label{eq:Lagrangian}
\end{equation}
where $L_B$ represents the Berry phase term, $L_B = (M_s/\gamma)\int dV \; \dot{\Phi}\left( 1 - \cos\Theta\right)$, and the Gilbert dissipation function is
\begin{equation}
W = \frac{1}{2}\alpha D \left( \dot{R}^2 + R^2 \dot{\Psi}^2 \right).
\end{equation}

We now expand this picture in order to account for scattering between the vortex core and magnons. We follow the collective-coordinate approach~\cite{rajaraman1982solitons} by expanding about the equilibrium profile, e.g., $\Theta(\mathbf{r},t) = \Theta_0[\mathbf{r}-\mathbf{X}(t)] + \eta[\mathbf{r}-\mathbf{X}(t),t]$, where the function $\eta$ describes spin wave fluctuations about the equilibrium state; a similar expansion about $\Phi(\mathbf{r},t)$ is also made. These fluctuations can be expressed in terms of the normal modes of the magnons, i.e., $\eta(\mathbf{r},t) \simeq ~ \sum_{n,m} a_{n,m}(t)\psi_{n,m}(\mathbf{r})$, where $a_{n,m}$ is the (complex) normal mode amplitude and $\psi_{n,m}(\mathbf{r})$ represents the spatial profile of this mode, such as those shown in the middle inset of Fig.~\ref{fig:SpectrProfil}. To second order in the fluctuations, it can be shown~\cite{heins2024self} that additional terms such as
\begin{equation}
L_B^{(2)} \sim \sum_{n,m}i \, C_{n,m}\left( a_{n,m}^* a_{n,m+1}V_g^{+} - a_{n,m} a_{n,m+1}^* V_g^{-}  \right)
\label{eq:intLB}
\end{equation}
appear in the Berry-phase term, where $C_{n,m}$ is a coupling constant and 
\begin{equation}
V_g^{\pm} = e^{\pm i \Psi(t)} \left( \dot{R} \pm i R \dot{\Psi}  \right),
\end{equation}
which can be interpreted as a velocity operator for the gyrotropic mode. Equation~(\ref{eq:intLB}) formalizes the three-particle scattering picture discussed previously, whereby we interpret the first term on the right-hand side as the annihilation of a $|n, m+1 \rangle$ mode and the creation of a $|n,m \rangle$ and $| g \rangle$ mode, while the second term represents the opposite process. The second order term $L_B^{(2)}$ therefore generates additional contributions to Eq.~(\ref{eq:ThielePolar}) by augmenting Eq.~(\ref{eq:Lagrangian}).

We now seek to develop a phenomenological scattering theory to account for Floquet magnons within an extension to Eq.~(\ref{eq:ThielePolar}). We will restrain the discussion to the Floquet comb modes, as shown in Fig.~\ref{fig:SpectrProfil}. Since the Floquet modes conserve the azimuthal symmetry of the linear modes, we will assume that scattering terms such as Eq.~(\ref{eq:intLB}) remain valid for the Floquet modes (albeit with different numerical values for the scattering amplitudes), so we can substitute $m$ for $\mu$. To simplify the notation and without loss of generalisation, we will take $n=0$ and use the comb index $k$, such that $k=0$ corresponds to the directly excited $\mu = -1$ mode at $f_\mathrm{rf} = 11.5$~GHz, i.e., $k=\mu + 1$. We can write the equation of motion of this directly excited mode as (with $\omega_\mathrm{rf} = 2 \pi f_\mathrm{rf}$)
\begin{equation}
\dot{a}_0 = -i \omega_\mathrm{rf} \, a_0 - \Gamma_0 a_0 + i P_0 b_\mathrm{rf} e^{-i \omega_\mathrm{rf} t},
\end{equation}
where $\Gamma_0$ is the linear relaxation rate and $P_0$ is a coupling constant with the external driving field. The steady-state solution to this inhomogeneous differential equation (neglecting transients) is
\begin{equation}
a_0(t) = i \left(\frac{P_0 b_\mathrm{rf}}{\Gamma_0} \right) e^{-i \omega_\mathrm{rf} t},
\label{eq:modeamp}
\end{equation}
and represents the source of the frequency comb. Consider the first adjacent comb modes to $a_0$, i.e., $a_{\pm1}$. These possess frequencies of $\omega_{\pm 1} = \omega_\mathrm{rf} \pm \omega_{g}$, with a steady state amplitude $\tilde{a}_{\pm 1}$ that is determined from nonlinear scattering with other magnons and the gyration through terms such as in Eq.~(\ref{eq:intLB}). Substituting these time dependencies into Eq.~(\ref{eq:intLB}) and evaluating Eq.~(\ref{eq:Lagrangian}), we find the modified equation of motion for $R$ to be
\begin{equation}
\dot{R} = - \Gamma_{g} R + 2 \left(\frac{P_0 b_\mathrm{rf}}{\Gamma_0} \right) \frac{1}{G \left(1 + \kappa^2 \right)}\left( C_{-1}\tilde{a}_{-1} - C_{1}\tilde{a}_{1} \right) \omega_g.
\label{eq:radialcomb}
\end{equation}
The second term on the right-hand side therefore represents an additional driving term for the radial dynamics, which scales with the driving field amplitude $b_\mathrm{rf}$ and the gyration frequency $\omega_g$, and is dictated by the balance between the occupation of the adjacent comb mode amplitudes $\tilde{a}_{\pm 1}$. As discussed elsewhere~\cite{heins2024self}, the Floquet mode amplitudes also depend on the gyration radius. We can generalize this expression to the $2N$ adjacent modes in the comb with the function $f(R)$, termed the nonlinear interaction contribution (NLC),
\begin{equation}
f(R) \equiv \frac{2 \omega_g }{G \left(1 + \kappa^2 \right)} \sum_{k=1}^{N} \left( C_{-k}\tilde{a}_{-k}\tilde{a}_{-k+1} - C_{k}\tilde{a}_{k}\tilde{a}_{k-1} \right),
\label{eq:NLC}
\end{equation}
such that
\begin{equation}
\dot{R} = - \Gamma_{g} R + f(R).
\label{eq:modelRg}
\end{equation}
Thus, depending on the functional form of $f(R)$, the dynamical system may admit nontrivial solutions for steady-state gyration, i.e., $\dot{R} = 0$ at $\Gamma_g R_0 = f(R_0)$, where $R_0 \neq 0$.

\subsection{Multiple stable gyration radii}
In this section, we discuss how simulated Floquet spectra can be used to determine $f(R)$ numerically. This can be achieved in the following way. First, starting from the equilibrium vortex ground state, we apply a rotating RF field $b_x + i b_y = b_r \exp(-i \omega_g t)$ in order to drive the core to steady state gyration. The rotating field strength $b_r$ determines the steady state radius $R_g$ once transients die out. Second, after steady-state is reached, an additional, low-amplitude linearly-polarized field $b_\mathrm{rf}$ along $\hat{\mathbf{x}}$ ($\approx 0.01$~mT), as considered previously, at a frequency $f_\mathrm{rf}$ is added to the rotating field. In contrast to the cases studied previously, we choose low field amplitudes here so as to not modify the gyration radius, but serving only to initiate the frequency comb around $f_\mathrm{rf}$. From the resulting power spectra of the Floquet modes (\emph{cf} Fig.~\ref{fig:SpectrProfil}), we obtain numerical estimates of $f(R)$ by using the five sidebands of the central frequency within the comb. We assume $f(R)$ to scale linearly with $b_\mathrm{rf}$, based on arguments leading to Eq.~(\ref{eq:radialcomb}). However, the functional form can vary considerably with $f_\mathrm{rf}$, as we will discuss next.

\begin{figure}
	\centering
	\includegraphics[width=8.5cm]{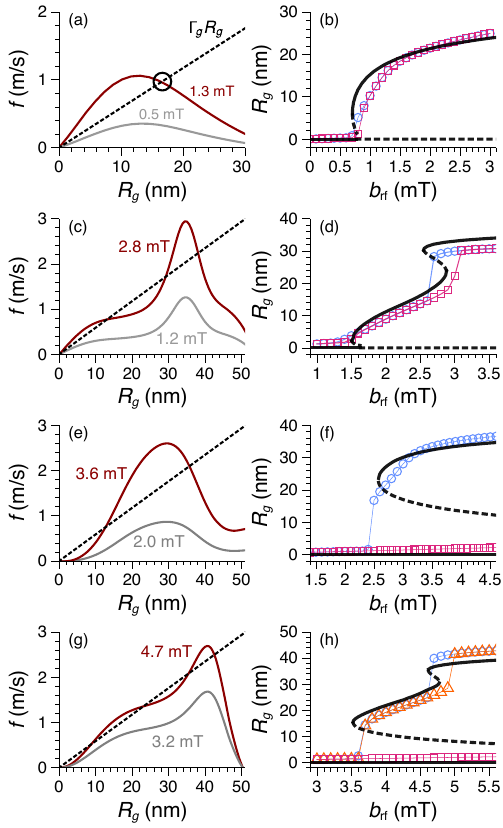}
	\caption{(a,c,e,g) Nonlinear interaction contribution to the gyration, $f(R_g)$, as a function of the steady-state gyration radius, $R_g$, for different driving frequencies, $f_\mathrm{rf}$: (a) 11.5~GHz, (c) 11~GHz, (e) 12.48~GHz, (g) 12.29~GHz. For each $f_\mathrm{rf}$, $f(R_g)$ is shown for two different values of $b_\mathrm{rf}$, below (grey) and above (red) threshold. The black, dashed line represents the relaxation function $\Gamma_g R_g$. The black circle in (a) indicates the point at which the relaxation is compensated by the nonlinear interaction contribution. (b,d,f,h) $R_g$ as a function of driving field amplitude, $b_\mathrm{rf}$, corresponding to the different $f_\mathrm{rf}$ in (a,c,e,g), i.e., $f_\mathrm{rf}$: (b) 11.5~GHz, (d) 11~GHz, (f) 12.48~GHz, (h) 12.29~GHz. Solid lines represent stable fixed points and dashed lines represent unstable fixed points, determined from Eq.~(\ref{eq:modelRg}). Open symbols represent the steady-state gyration radius extracted from simulations after an interval of 200~ns, with an initial state with gyration radius of 0~nm (squares), 41~nm (circles), and 19~nm (triangles).}
	\label{fig:RadiusCore}
\end{figure}

An example of $f(R)$ extracted from simulations is shown in Fig.~\ref{fig:RadiusCore}(a), corresponding to the case of $f_\mathrm{rf}=11.5$~GHz.
$f(R)$ is a concave function over the range of $R_g$ considered and is plotted as solid lines for two values of $b_\mathrm{rf}$. The linear, dotted curve in Fig.~\ref{fig:RadiusCore}(a) represents the relaxation function $\Gamma_g R$. With this representation, we can determine pictorially the fixed points of the radial dynamics. At $b_\mathrm{rf}=0.5$~mT, $f(R)$ and $\Gamma_g R_g$ only intersect at $R_g^* = 0$, with the damping rate exceeding the NLC everywhere, which indicates that no steady-state gyration can be induced by the high-frequency excitation. In this case, $R_g^* = 0$ is a stable fixed point. For a higher field of $b_\mathrm{rf}=1.3$~mT, on the other hand, the two curves intersect at the origin and at $R_g^* \approx 16.5$~nm (as illustrated by the circle in Fig.~\ref{fig:RadiusCore}(a)), indicating the presence of two fixed points. At this higher field, $R_g^* = 0$ becomes an unstable fixed point, since the NLC exceeds the damping for any finite excursion from the disk center, and $R_g^* \approx 16.5$~nm is a stable fixed point, indicating that steady-state gyration is possible at this driving field.

By evaluating the fixed points $R_g^*$ for a range of $b_\mathrm{rf}$, we can predict the threshold field and subsequent variation of the steady-state gyration radius. This result is presented in Fig.~\ref{fig:RadiusCore}(b) for $f_\mathrm{rf}=11.5$~GHz, based on the function $f(R)$ in Fig.~\ref{fig:RadiusCore}(a). Predictions from the model are shown as black lines, where solid lines indicate a stable fixed point, while dashed lines indicate an unstable fixed point. We can observe a field threshold at around $b_\mathrm{rf} = 0.75$~mT, consistent with the result found earlier in Fig.~\ref{fig:FreqComb}(d). Above this threshold, the stable gyration orbit $R_g$ grows with $b_\mathrm{rf}$, while the origin $R_g^* = 0$ remains unstable. We tested the stability of these predicted fixed points in the following way. For each $b_\mathrm{rf}$ considered, we run simulations over an interval of 200 ns in which the initial radial core position is either 0 nm (squares) or 41 nm (circles). The gyration radius at the end of the simulation is then recorded and shown in Fig.~\ref{fig:RadiusCore}(b). We can observe that both initial states end up at the predicted stable fixed point with relatively good quantitative agreement, with discrepancies only seen around the threshold. The variation in Fig.~\ref{fig:RadiusCore}(b) is consistent with a supercritical Hopf bifurcation, as observed in spin-torque vortex oscillators when an applied dc current is increased past a threshold~\cite{ivanov2007excitation, kim2012spin}. We can therefore interpret the dynamics here as a Floquet-driven auto-oscillation of the vortex core.

Other kinds of bifurcations are observed, leading to multiple stable gyration radii and possible hysteretic effects. An example is presented in Fig.~\ref{fig:RadiusCore}(c) for $f_\mathrm{rf} = 11$~GHz. Above threshold at $b_\mathrm{rf} = 2.8$~mT, we see that the relaxation curve intersects the NLC function at four points, resulting in two stable and and two unstable fixed points. The evolution of these fixed points is presented in Fig~\ref{fig:RadiusCore}(d), where we observe a first threshold at around $b_\mathrm{rf} \approx 1.5$~mT, giving rise to a single, finite steady-state gyration radius, followed by a second bifurcation at $b_\mathrm{rf} \approx 2.5$~mT. Within the approximate field interval of 2.5 to 2.9~mT, we observe the presence of the four fixed points seen in Fig.~\ref{fig:RadiusCore}(c). Simulations suggest that hysteretic effects can appear at this driving frequency, since the stable fixed point occupied depends on initial conditions; initial states at low radii evolve toward the square symbols, while higher radii evolve toward the circle symbols, with a clear opening between the two curves in the interval between 2.5 and 2.9~mT. This means that upward and downward $b_\mathrm{rf}$ sweeps would become hysteretic, as seen in recent experiments~\cite{heins2024self}. For a higher driving frequency of $f_\mathrm{rf} = 12.48$~GHz, the form of $f(R)$ [Fig.~\ref{fig:RadiusCore}(e)] results in a saddle-node bifurcation at around $b_\mathrm{rf} \approx 2.6$~mT in which the origin remains a stable fixed point, but with a pair of stable and unstable fixed points emerging from $R_g \approx 23$~nm [Fig.~\ref{fig:RadiusCore}(f)]. Simulations show that hysteretic behavior can appear over a wide range of $b_\mathrm{rf}$. Finally, driving at $f_\mathrm{rf} = 12.29$~GHz results in a large number of fixed points, as seen by the form of $f(R)$ in Fig.~\ref{fig:RadiusCore}(g). The model predicts that up to three stable fixed points can be present simultaneously, arising from a sequence of bifurcations as shown in Fig.~\ref{fig:RadiusCore}(h). Here, a third initial condition was considered in the simulations, i.e., an initial gyration radius of 19~nm (triangles). The functional form of $f(R)$ is therefore strongly dependent on the driving frequency and underpins a rich variety of possible fixed points and associated bifurcations. Moreover, good quantitative agreement is found overall between the predicted fixed points and simulated gyration radii.

\begin{figure}
	\centering
	\includegraphics[width=8.5cm]{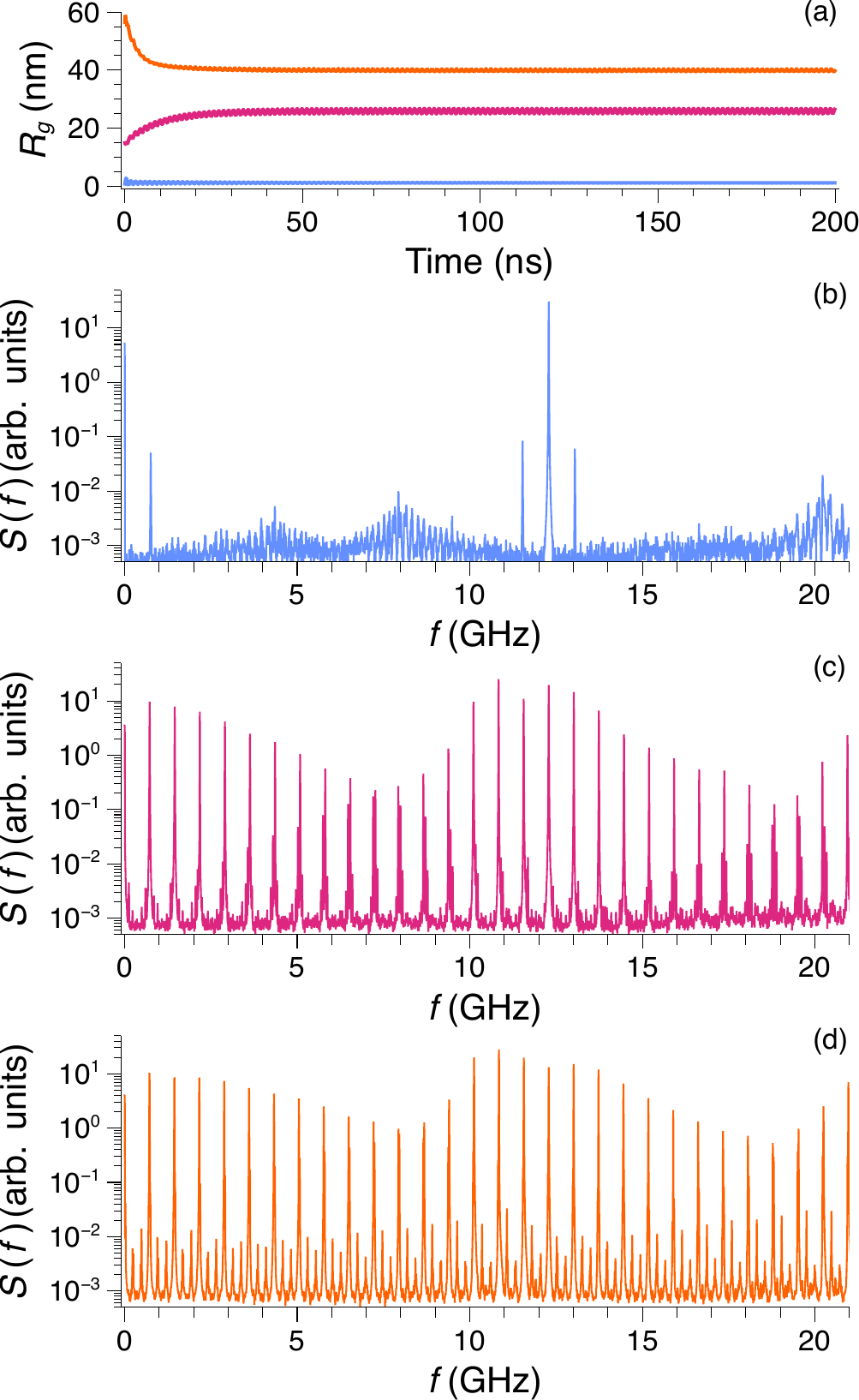}
	\caption{(a) Time evolution of the gyration radius for $f_\mathrm{rf} = 12.29$~GHz and $b_\mathrm{rf}=4.7$~mT, resulting from three different simulations with initial conditions for the gyration radius. (b,c,d) Power spectral density, $S(f)$, for three different average steady-state gyration radii, $R_g$: (b) $\approx 1$~nm, (c) $\approx 26$~nm, (d) $\approx 40$~nm.}
	\label{fig:SPECTRA}
\end{figure}

Since the Floquet spectra should be dependent on the gyration radius, we highlight an example in which a single high-frequency drive can result in different spectral response by virtue of the multiple stable $R_g^*$. We revisit the $f_\mathrm{rf} = 12.29$~GHz case above, where three distinct $R_g^*$ can be found at $b_\mathrm{rf} = 4.7$~mT. The resulting power spectral density is presented in Fig.~\ref{fig:SPECTRA}.
For this driving frequency, we conducted three separate simulations in which the initial state comprised distinct gyration radii, as shown in Fig.~\ref{fig:SPECTRA}(a). As predicted from the model, the system settles into three different steady-state gyration radii, consistent with the results in Fig.~\ref{fig:RadiusCore}(g). Once the gyration is stable, we ran the simulations over 200~ns and computed the PSD, as in Fig.~\ref{fig:SpectrProfil}. For the lowest gyration radius considered [Fig.~\ref{fig:SPECTRA}(b)], we observe primarily a single response to the driving field at $f_{rf} = 12.29$~GHz, with a much weak pair of sidebands associated with the gyration. For the moderate [Fig.~\ref{fig:SPECTRA}(c)] and highest [Fig.~\ref{fig:SPECTRA}(d)] gyration radii, a rich frequency comb is observed, corresponding to the Floquet magnons induced by the gyration and the direct field pumping. Interestingly, the form of the frequency comb contain distinct differences between the two cases, which suggests that the comb spectra might be useful probe for detecting hysteretic behavior related to transitions between multiple gyration radii.

%%
%	Section: Discussion and concluding remarks
%%
\section{Discussion and concluding remarks}
We have presented a detailed theoretical and computational study on how the excitation of azimuthal spin waves in vortex-state nanopillars can drive the vortex core into steady gyration, resulting in the appearance of frequency combs related to the emergence of magnon Floquet modes. This is termed a ``self-induced'' mechanism~\cite{heins2024self} for generating Floquet modes, since the gyration is not excited directly by the external drive but rather through nonlinear interactions between the directly-excited azimuthal modes and the vortex core. The fact that the same mechanism is observed within the 300~nm-diameter, 20~nm-thick disks we study and in thicker, larger 5 $\mu$m-diameter disks reported elsewhere~\cite{heins2024self} suggests that the mechanism is common to all vortex-based systems, with the necessary driving frequency depending only on the corresponding frequencies of the lowest-order azimuthal modes. While discussion has been focused on $n=0$ modes, we also have evidence from simulations that the self-induced mechanism is also present for the $n=1$ azimuthal modes, but occurs over a smaller interval of the driving frequency.

Based on a collective-coordinates approach, we have extended the Thiele model to account for how interactions between the core and Floquet magnons can sustain core-gyration. This involved including three-particle scattering terms in the Lagrangian, which describe the confluence and splitting of azimuthal modes with the core gyration. These scattering terms result in a nonlinear contribution to the radial dynamics of core. By parameterizing the model parameters with simulated Floquet spectra at different steady-state gyration radii, we can accurately predict the existence of multiple stable and unstable gyration radii under azimuthal mode pumping, with good quantitative agreement found with micromagnetics simulations. While we did not use the Floquet modes as a basis for deriving the scattering terms, the theory relies on Floquet modes preserving the same symmetries as the linear modes (i.e., azimuthal indices), so any differences relate only to scattering coefficients which we use as fitting parameters for the simulated Floquet spectra. The theory also relies on the assumption that steady-state gyration is reached. It may be worthwhile to extend this formalism in future work in order to account for the transient dynamics of the vortex core, e.g., how it is initially driven away from its equilibrium position, by including bilinear terms in the Lagrangian. From a simulation perspective, the transient dynamics of the relevant magnon modes could also be examined using a mode-resolved approach~\cite{massouras2024mode}.

We suggest that Floquet states resulting in multiple stable gyration radii could be observed experimentally, by measuring the power spectra as the driving field amplitude, frequency, or both are swept back and forth. Because the Floquet spectra acquire a dependence on the driving field history, one might envisage interesting extensions to recently proposed unconventional computing schemes based on nonlinear magnon scattering~\cite{koerber2023pattern, heins2025benchmarking}. In that work, it was shown that the transient dynamics related to magnon scattering allow pattern recognition tasks involving RF pulse sequences to be performed, where inputs involve directly-excited radial modes $(n,m=0)$ and outputs involve scattered azimuthal modes ($n',m\neq 0$). With our system, a fruitful avenue of exploration might involve exploring whether the entire frequency comb could serve as a useful output, since its structure appears to be greatly dependent on the drive frequency, amplitude, and history. Along these lines, interesting parallels may be found with recent proposals for photonic reservoir computing~\cite{butschek2022photonic}, where frequency-multiplexing using frequency combs has been studied for challenging computational tasks like chaotic time series prediction.

\section*{acknowledgements}
We thank Lukas K\"{o}rber for fruitful discussions and a critical reading of the manuscript. This work was supported by the European Commission's Research and Innovation Programme Horizon Europe under grant agreement no. 101070290 (NIMFEIA).

\bibliography{bibliography}

\end{document}